\documentclass[11pt,a4paper]{article}

\usepackage{amsmath,amsthm,latexsym,amssymb}
\usepackage{graphicx,bm,fancybox,multirow,hhline,indentfirst}
\usepackage{ascmac,atbegshi}
\usepackage{natbib,verbatim,color}
\usepackage[colorlinks,citecolor=blue,urlcolor=blue,dvipdfmx]{hyperref}

\makeatletter
\def\section{\@startsection {section}{1}{\z@}{-3.5ex plus -1ex minus -.2ex}{2.3 ex plus .2ex}{\large\sc\centering}}
\def\subsection{\@startsection {subsection}{1}{\z@}{-3.5ex plus -1ex minus -.2ex}{2.3 ex plus .2ex}{\large}}
\makeatother

\theoremstyle{definition}
\newtheorem{theorem}{Theorem}

\title{\Large\bf $C_p$ criterion for semiparametric approach in causal inference\medskip
}
\author{\large Takamichi Baba\thanks{1-1-4 Shibata, Kita-ku, Osaka 530-0012, Japan. Email: takamichi.baba@shionogi.co.jp}\\{\it Biostatistics Department, Shionogi \& Co., Ltd.}  \medskip \and Yoshiyuki Ninomiya\thanks{Corresponding author. 744 Motooka, Fukuoka 819-0395, Japan. Email: nino@imi.kyushu-u.ac.jp}\\{\it  Institute of Mathematics for Industry, Kyushu University} \medskip}
\date{\normalsize Version: \today}

\renewcommand{\baselinestretch}{1.6}\selectfont

\allowdisplaybreaks

\def\QIC{{\rm QIC}}
\def\MuSE{{\rm MuSE}}
\def\MwSE{{\rm MwSE}}
\def\wcp{\mbox{{\rm w}$C_p$}}
\def\ucp{\mbox{{\rm u}$C_p$}}
\def\T{{\mathrm{\scriptscriptstyle T}}}
\def\E{{\rm E}}
\def\P{{\rm P}}
\def\oP{{\rm o}_{\rm P}}

\begin{document}

\maketitle

\begin{abstract}
\noindent
For marginal structural models, which recently play an important role in causal inference, we consider a model selection problem in the framework of a semiparametric approach using inverse-probability-weighted estimation or doubly robust estimation.
 In this framework, the modeling target is a potential outcome which may be a missing value, and so we cannot apply the AIC nor its extended version to this problem.
 In other words, there is no analytical information criterion obtained according to its classical derivation for this problem.
 Hence, we define a mean squared error appropriate for treating the potential outcome, and then we derive its asymptotic unbiased estimator as a $C_{p}$ criterion from an asymptotics for the semiparametric approach and using an ignorable treatment assignment condition.
 In simulation study, it is shown that the proposed criterion exceeds a conventionally derived existing criterion in the squared error and model selection frequency.
 Specifically, in all simulation settings, the proposed criterion provides clearly smaller squared errors and higher frequencies selecting the true or nearly true model. 
 Moreover, in real data analysis, we check that there is a clear difference between the selections by the two criteria. 

\

\end{abstract}

\newpage

\section{Introduction}\label{sec1}

The marginal structural model  (\citealt{Rob97}, \citealt{RobHer00}) is one of the most basic models in causal inference.
 This is a potential outcome model, and the data are regarded to be partly missed.
 Therefore, if we do estimation naively despite that the outcome and missing mechanism are correlated, the estimator will have a large bias.
 While this bias is removed if we can correctly specify the correlation, it is common to rely on a semiparametric approach using inverse-probability-weighted estimation (\citealt{RobRZ94}) or doubly robust estimation (\citealt{SchRR99}, \citealt{BanRob05}) without the difficult modeling.

As an example, let us consider a simple marginal structural model $y_i^{(h)}=\sum_{j=0}^pb_{j+1}x^{(h)j}+\varepsilon_i$ (\citealt{PlaBro13}, \citealt{TalARBL15}), where $y_i^{(h)}$ is a potential outcome for the $i$-th sample with the treatment $x^{(h)}$, $t_i^{(h)}$ is an indicator which is $1$ if the treatment $x^{(h)}$ is received and $0$ otherwise, and $\varepsilon_i$ is an error.
 In this model, $y_i^{(h)}$ with $t_i^{(h)}=0$ is regarded as being missed.
 Therefore, if we estimate the regression form $\sum_{j=0}^pb_{j+1}x^{(h)j}$ by the least squares method in spite of existing the correlation between $y_i^{(h)}$ and $t_i^{(h)}$, a bias yields as a matter of course.
 Then, supposing that a confounder $\bm{z}_i$ between $y_i^{(h)}$ and $t_i^{(h)}$ is observed, a semiparametric approach using the propensity score $e_i^{(h)}\equiv\P(t_i^{(h)}=1\mid\bm{z}_i)$  (\citealt{RosRub83}) is commonly used.
Under this setting, we treat a model selection problem for the regression form of interest, which is the selection problem of the order $p$ in the polynomial in this example. 

To be surprising, there is no information criterion made by adjusting classical ones to this basic problem except for one. 
 The valuable one is \QIC$_w$ in \cite{PlaBro13}. 
 This criterion is made by replacing the goodness-of-fit term in \QIC\ (\citealt{Pan01}), the quasi-maximum log-likelihood, with a quasi-maximum weighted log-likelihood in order to cope with the missing values.
 That is, \QIC$_w$ uses the same penalty term as in \QIC\ although \QIC\ does not cope with the missing values.
 In this paper, we show that if we evaluate the penalty term based on the original definition of an information criterion, it becomes quite different term from \QIC's. 

As written in \cite{PlaBro13}, while the model selection problem for the regression form is little treated, the confounder selection problem is treated in, for example, \cite{BroLaa06} and \cite{VanBC12}.
 These papers use a cross-validation-type method with a high computational cost or the FIC (\citealt{ClaHjo03}) based on a special assumption of local misspecification. 
 In this paper, it is not considered to develop them for our problem, and we construct a method without relying on such a computational cost or special assumption. 

In Section \ref{sec2}, the model and assumption are explained, and we introduce the inverse-probability-weighted estimation and doubly robust estimation under them.
 In Section \ref{sec3}, first we give two kinds of mean squared errors, \MwSE\ and \MuSE, appropriate for treating missing mechanism, and then we get goodness-of-fit and penalty terms similarly to in the derivation of the conventional $C_p$ criterion.
 Note that the goodness-of-fit term in \MwSE\ becomes the same one as in \QIC$_w$.
 Next, we asymptotically evaluate the penalty terms for the inverse-probability-weighted and doubly robust estimations by using techniques similar to in showing the consistency of these estimators. 
 As a result, this asymptotic $C_p$ takes the form we can easily evaluate, and we set it as our proposed criterion. 
 In Sections \ref{sec4} and \ref{sec5}, we compare the performances of the existing and proposed criteria through simulation studies under basic situations as mentioned above and real data analysis, respectively.
In Section \ref{sec6}, to explore the possibility for improvement and generalization of the proposed criterion, we mention about modifying the mean squared error and applying it to missing data analysis.


\section{Preliminary}\label{sec2}

\subsection{Model and assumptions}\label{sec2_1}

The marginal structural mean model is a model for the marginal means of potential outcomes.
 Let us assume that there are $H$ kinds of treatments, and we denote a potential outcome for the $h$-th treatment by $\bm{y}^{(h)}\ (\in\mathbb{R}^m)$, and let $t^{(h)}$ be a random indicator which is $1$ if the $h$-th treatment is received and $0$ otherwise ($h\in\{1,2,\ldots,H\},\ \sum_{h=1}^Ht^{(h)}=1$).
 Then, we consider a marginal structural model 
\begin{align*}
\bm{y} = \sum_{h=1}^{H} t^{(h)} \bm{y}^{(h)} = \sum_{h=1}^{H} t^{(h)} \left( \bm{X}^{(h)} \bm{\beta} + \bm{\varepsilon} \right),
\end{align*}
which assumes a linear regression model by each potential outcome.
 In the right hand side, $\bm{X}^{(h)}\ (\in\mathbb{R}^{m\times p})$ is an independent variable matrix, $\bm{\varepsilon}\ (\in\mathbb{R}^m)$ is an error vector whose mean is $\bm{0}$ and dispersion matrix is $\sigma^2\bm{I}_m$, where $\bm{0}$ is a zero vector or a zero matrix and $\bm{I}$ is an identity matrix.
 Note that $\bm{y}$ in the left hand side is an observed outcome. 
 In this model, $H-1$ potential outcomes, $\bm{y}^{(h)}$'s with $t^{(h)}=0$, are regarded as missing values.
 Therefore, if we estimate $\bm{X}^{(h)}\bm{\beta}$ naively from observed outcomes, the estimator will have a bias because $\E[\bm{y}^{(h)}]\neq \E[\bm{y}^{(h)}\mid t^{(h)}=1]$ in general.
 In this paper, we suppose that a confounder vector $\bm{z}\ (\in\mathbb{R}^s)$ between $\bm{y}^{(h)}$ and $t^{(h)}$ is observed so that this bias can be removed.

For this model, we make several basic assumptions.
 First, let us consider $\bm{X}^{(h)}$.
 Although we consider a non-random variable as the components of $\bm{X}^{(h)}$ in the example in Section \ref{sec1}, here we allow it to include a part of confounder vector $\bm{z}$ in order to treat more general setting.
 In addition, to reduce the complexity of expressions, we assume that these independent variables are standardized so that $\E[\sum_{h=1}^{H}\bm{X}^{(h)\T}\bm{X}^{(h)}]=\bm{I}$.
 This assumption is not essential, and actually the final form of the derived criterion in the following does not depend on whether we make this assumption or not.
 Next, we assume a weakly ignorable treatment assignment condition (\citealt{Imb00})
\begin{align*}
\bm{y}^{(h)} \perp t^{(h)} \mid \bm{z} \hspace{10mm} (h \in \{1,2,\ldots,H\}),
\end{align*}
which is to assure that we can remove the above-mentioned bias.
 Note that we can replace $\bm{y}^{(h)}$ with $\bm{\varepsilon}$ in this condition.

Now we have $N$ samples following this model, and we put subindex $i$ in variables for the $i$-th sample.
 In addition, let $\tilde{\bm{y}}^{(h)}=(\bm{y}_{1}^{(h)\T},\bm{y}_{2}^{(h)\T},\ldots,\bm{y}_{N}^{(h)\T})^{\T}$, $\bm{T}^{(h)}={\rm diag}(t_{i}^{(h)}\bm{I})$, $\tilde{\bm{X}}^{(h)}=(\bm{X}_{1}^{(h)\T},\bm{X}_{2}^{(h)\T},\ldots,\bm{X}_{N}^{(h)\T})^{\T}$ and $\tilde{\bm{\varepsilon}}=(\bm{\varepsilon}_{1}^{\T},\bm{\varepsilon}_{2}^{\T},\ldots,\bm{\varepsilon}_{N}^{\T})^{\T}$, and then we can express the model by
\begin{align*}
\tilde{\bm{y}} = \sum_{h=1}^{H} \bm{T}^{(h)} \tilde{\bm{y}}^{(h)} = \sum_{h=1}^{H} \bm{T}^{(h)} \left( \tilde{\bm{X}}^{(h)} \bm{\beta} + \tilde{\bm{\varepsilon}} \right).
\end{align*}
Here, we assume that the samples are independent each other, that is, 
\begin{align*}
(t_{i}^{(h)},\bm{X}_{i}^{(h)},\bm{\varepsilon}_{i},\bm{z}_{i}) \perp (t_{j}^{(h)},\bm{X}_{j}^{(h)},\bm{\varepsilon}_{j},\bm{z}_{j}) \hspace{10mm} (i\neq j, \ h\in\{1,2,\ldots,H\}).
\end{align*}
From this, it holds $\bm{y}_{i} \perp \bm{y}_{j}$ $(i \neq j)$ as a matter of course.
Moreover, we assume that $\bm{X}_i^{(h)}$ and $\bm{\varepsilon}_i$ are independent as done for conventional regression models.

\subsection{Estimation method}\label{sec2_2}

If the relationship between the potential outcome $\bm{y}^{(h)}$ and confounder $\bm{z}$ is correctly modeled, we can easily give a consistent estimator of the marginal mean for $\bm{y}^{(h)}$ under the ignorable treatment assignment condition.
 However, this modeling is difficult in general.
 Therefore, in recent years, it is often the case that we rely on a semiparametric approach using so-called the propensity score, $e_{i}^{(h)} (\bm{\alpha}) \equiv \P (t_{i}^{(h)}=1 \mid \bm{z}_i; \bm{\alpha})$, which does not depend on the correct modeling. 
 Here, $\bm{\alpha}\ (\in\mathbb{R}^q)$ is a parameter vector relating to the propensity score.
 In this paper, we treat two kinds of estimation methods basic in this approach.

The first one is the inverse-probability-weighted estimation (\citealt{RobRZ94}).
In this method, missing values are restored through weighting the observed values by the inverse of the propensity score, and then a conventional estimation is used. 
Specifically, we define a weighted squared loss function as 
\begin{align}
\sum_{h=1}^{H} \left( \tilde{\bm{y}} - \tilde{\bm{X}}^{(h)} \bm{\beta} \right)^{\T} \bm{W}^{(h)} (\bm{\alpha}) \left( \tilde{\bm{y}} - \tilde{\bm{X}}^{(h)} \bm{\beta} \right)
\label{IPWloss}
\end{align}
using a weight matrix $\bm{W}^{(h)}(\bm{\alpha})\equiv{\rm diag}\{t^{(h)}_{i}\bm{I}_{r}/e_{i}^{(h)}(\bm{\alpha})\}$, and then the inverse-probability-weighted estimator
\begin{align}
\hat{\bm{\beta}}^{\rm IPW} (\bm{\alpha}) \equiv \left\{ \sum_{h=1}^{H} \tilde{\bm{X}}^{(h)\T} \bm{W}^{(h)}(\bm{\alpha}) \tilde{\bm{X}}^{(h)} \right\}^{-1} \sum_{h=1}^{H} \tilde{\bm{X}}^{(h)\T} \bm{W}^{(h)}(\bm{\alpha}) \tilde{\bm{y}}
\label{IPWest}
\end{align}
is given by minimizing the loss function with respect to $\bm{\beta}$.
 If $\bm{\alpha}$ is unknown, we obtain the maximum likelihood estimator $\hat{\bm{\alpha}}$ through $p(\tilde{\bm{t}} \mid \tilde{\bm{z}}; \bm{\alpha})$, the conditional probability function of $\tilde{\bm{t}} = (\bm{t}_1^{\T}, \bm{t}_2^{\T}, \ldots, \bm{t}_N^{\T})^{\T}$ given $\tilde{\bm{z}} = (\bm{z}_1^{\T}, \bm{z}_2^{\T}, \ldots, \bm{z}_N^{\T})^{\T}$, and we use it in place of $\bm{\alpha}$, where $\bm{t}_i=(t_i^{(1)}, t_i^{(2)}, \ldots, t_i^{(H)})^{\T}$.
 This inverse-probability-weighted estimator is consistent under the ignorable treatment assignment condition.

While $\bm{y}^{(h)}$ is correlated with $\bm{z}$ in general in the marginal structural model, the inverse-probability-weighted estimation does not directly use the information of $\bm{z}$ for estimating the marginal mean of $\bm{y}^{(h)}$. 
The doubly robust estimation (\citealt{SchRR99}, \citealt{BanRob05}) implements it to improve the inverse-probability-weighted estimation, and it uses $f(\tilde{\tilde{\bm{y}}} \mid \tilde{\bm{z}}; \bm{\gamma})$, the conditional probability density function of $\tilde{\tilde{\bm{y}}}=(\tilde{\bm{y}}^{(1)\T}, \tilde{\bm{y}}^{(2)\T}, \ldots, \tilde{\bm{y}}^{(H)\T})^{\T}$ given $\tilde{\bm{z}}$.
 Here, $\bm{\gamma}\ (\in\mathbb{R}^r)$ is a parameter vector relating to the conditional distribution.
 Denoting the expectation based on this conditional distribution by $\E[\cdot\mid\tilde{\bm{z}};\bm{\gamma}]$, the doubly robust estimator is given by minimizing with respcet to $\bm{\beta}$ the expression which is made by adding 
\begin{align*}
\sum_{h=1}^{H}
\left( \E \left[ \tilde{\bm{y}}^{(h)} \mid \tilde{\bm{z}}; \bm{\gamma} \right] - \tilde{\bm{X}}^{(h)} \bm{\beta}\right)^{\T} \left\{ \bm{I}-\bm{W}^{(h)}(\bm{\alpha}) \right\} \left( \E \left[ \tilde{\bm{y}}^{(h)} \mid \tilde{\bm{z}}; \bm{\gamma} \right] - \tilde{\bm{X}}^{(h)}\bm{\beta} \right)
\end{align*}
to \eqref{IPWloss}.
 In the framework of the doubly robust estimation, usually $\bm{\alpha}$ and $\bm{\gamma}$ are unknown, and so we replace them with the maximum likelihood estimators $\hat{\bm{\alpha}}$ and $\hat{\bm{\gamma}}$ which are obtained through $p(\tilde{\bm{t}} \mid \tilde{\bm{z}}; \bm{\alpha})$ and $f(\tilde{\tilde{\bm{y}}} \mid \tilde{\bm{z}}; \bm{\gamma})$, respectively.
 To avoid complex statements, hereafter we omit these arguments.
 Then, the doubly robust estimator is expressed as
\begin{align*}
\hat{\bm{\beta}}^{\rm DR} \equiv \left( \sum_{h=1}^{H} \tilde{\bm{X}}^{(h)\T} \tilde{\bm{X}}^{(h)}\right)^{-1} \sum_{h=1}^{H} \left\{ \tilde{\bm{X}}^{(h)\T} \bm{W}^{(h)} \tilde{\bm{y}} + \tilde{\bm{X}}^{(h)\T} \left( \bm{I}-\bm{W}^{(h)} \right) \E \left[ \tilde{\bm{y}}^{(h)} \mid \tilde{\bm{z}} \right] \right\}.
\end{align*}
This estimator not only improves the inverse-probability-weighted estimator but also achieves to be semiparametrically efficient (\citealt{RobRot95}). 
 In addition, when either the propensity score or the conditional distribution is correctly specified, the estimator is consistent.


\section{Proposed model selection criteria}\label{sec3}

\subsection{Mean squared errors for causal inference}\label{sec3_1}

Before defining a mean squared error for causal inference, we will explain about \QIC$_w$ proposed by \citet{PlaBro13}.
 When there are no missing data and the dispersion matrix of $\bm{\varepsilon}$ is $\sigma^2\bm{I}$, the criterion in \cite{Pan01} is written as
\begin{align*}
\QIC = \sum_{h=1}^H \left( \tilde{\bm{y}}^{(h)} - \tilde{\bm{X}}^{(h)} \hat{\bm{\beta}} \right)^{\T} \left( \tilde{\bm{y}}^{(h)} - \tilde{\bm{X}}^{(h)} \hat{\bm{\beta}} \right) +2\sigma^2p,
\end{align*}
where $\hat{\bm{\beta}}$ is a quasi-maximum likelihood estimator.
 This is an unbiased estimator of so-called a quasi-likelihood version of the Kullback-Leibler divergence, in other words, this is a $C_p$ criterion derived from the conventional mean squared error, and so \QIC\ is regarded as a reasonable criterion.
 On the other hand, when there are missing data, \QIC\ cannot be obtained and
\begin{align*}
\QIC_w = \sum_{h=1}^{H} \left( \tilde{\bm{y}}^{(h)} - \tilde{\bm{X}}^{(h)} \hat{\bm{\beta}} \right)^{\T} \bm{W}^{(h)} \left( \tilde{\bm{y}}^{(h)} - \tilde{\bm{X}}^{(h)} \hat{\bm{\beta}} \right) +2\sigma^{2}p
\end{align*}
is proposed.
 This criterion is based on the fact that if $\hat{\bm{\beta}}$ is the above-mentioned quasi-maximum likelihood estimator, it holds $\E[\QIC]=\E[\QIC_w]$ under the ignorable treatment assignment condition because $\E[\bm{W}^{(h)}\mid\tilde{\bm{z}}]=\bm{I}$.
 However, if $\hat{\bm{\beta}}$ is the inverse-probability-weighted estimator or the doubly robust estimator, it is not conditionally independent of $\bm{W}^{(h)}$, and so we have $\E[\QIC] \neq \E[\QIC_w]$ in general.
 Even more important is that $2\sigma^2p$ is a penalty for an estimator ignoring the existence of missing data and not for the semiparmetric estimator, and it becomes a problem in using \QIC$_w$.
 Actually, the variance of the latter estimator is much larger than that of the former estimator, and so we need to enlarge the penalty for the latter estimator.

Hence, let us consider two kinds of appropriate mean squared errors for the case where there are missing data.
As the first kind, we define a mean weighted squared error by
\begin{align}
\MwSE = & \sum_{h=1}^{H} \E \left[ \left(  \tilde{\bm{X}}^{(h)} \hat{\bm{\beta}} - \E \left[ \tilde{\bm{y}}^{(h)} \mid \tilde{\bm{X}}^{(h)} \right] \right)^{\T} \bm{W}^{(h)} \left( \tilde{\bm{X}}^{(h)} \hat{\bm{\beta}} - \E \left[ \tilde{\bm{y}}^{(h)} \mid \tilde{\bm{X}}^{(h)} \right] \right) \right] 
\nonumber \\
= & \sum_{h=1}^{H} \E \left[ \left( \tilde{\bm{y}} - \tilde{\bm{X}}^{(h)} \hat{\bm{\beta}} \right)^{\T} \bm{W}^{(h)} \left( \tilde{\bm{y}} - \tilde{\bm{X}}^{(h)} \hat{\bm{\beta }} \right) \right] 
\nonumber \\&
- \sum_{h=1}^{H} \E \left[ \left( \tilde{\bm{y}} - \E \left[ \tilde{\bm{y}}^{(h)} \mid \tilde{\bm{X}}^{(h)} \right] \right)^{\T} \bm{W}^{(h)} \left( \tilde{\bm{y}} - \E \left[ \tilde{\bm{y}}^{(h)} \mid \tilde{\bm{X}}^{(h)} \right] \right) \right] \nonumber \\
& + 2\sum_{h=1}^{H} \E \left[ \left( \tilde{\bm{y}} - \E \left[ \tilde{\bm{y}}^{(h)} \mid \tilde{\bm{X}}^{(h)} \right] \right)^{\T} \bm{W}^{(h)} \left( \tilde{\bm{X}}^{(h)} \hat{\bm{\beta}} - \E \left[ \tilde{\bm{y}}^{(h)} \mid \tilde{\bm{X}}^{(h)} \right] \right) \right] .
\label{WSE}
\end{align}
According to the derivation of the conventional $C_p$ criterion, this is decomposed into three terms after the definition.
 This sum of weighted squared differences can be regard as the sum of squared differences between the expectations for the data restored by using the weight, which is also used in the inverse-probability-weighted estimation, and their estimators.
 Actually, the first term in the decomposition is the expectation of \eqref{IPWloss}.
 That is, we consider the same loss function in the derivation and in the error evaluation for the estimator, and so it is natural in that term.
 As the second kind, we define a mean unweighted squared error by
\begin{align}
\MuSE = & \sum_{h=1}^{H} \E \left[ \left( \tilde{\bm{X}}^{(h)} \hat{\bm{\beta}} - \E \left[ \tilde{\bm{y}}^{(h)} \mid \tilde{\bm{X}}^{(h)} \right] \right)^{\T} \bm{T}^{(h)} \left( \tilde{\bm{X}}^{(h)} \hat{\bm{\beta}} - \E \left[ \tilde{\bm{y}}^{(h)} \mid \tilde{\bm{X}}^{(h)} \right] \right) \right] \nonumber \\
= & \sum_{h=1}^{H} \E \left[ \left( \tilde{\bm{y}} - \tilde{\bm{X}}^{(h)} \hat{\bm{\beta}} \right)^{\T} \bm{T}^{(h)} \left( \tilde{\bm{y}} - \tilde{\bm{X}}^{(h)} \hat{\bm{\beta}} \right) \right] \nonumber \\
& - \sum_{h=1}^{H} \E \left[ \left( \tilde{\bm{y}} - \E \left[ \tilde{\bm{y}}^{(h)} \mid \tilde{\bm{X}}^{(h)} \right] \right)^{\T} \bm{T}^{(h)} \left( \tilde{\bm{y}} - \E \left[ \tilde{\bm{y}}^{(h)} \mid \tilde{\bm{X}}^{(h)} \right] \right) \right] \nonumber \\
& + 2\sum_{h=1}^{H} \E \left[ \left( \tilde{\bm{y}} - \E \left[ \tilde{\bm{y}}^{(h)} \mid \tilde{\bm{X}}^{(h)} \right] \right)^{\T} \bm{T}^{(h)} \left( \tilde{\bm{X}}^{(h)} \hat{\bm{\beta}} - \E \left[ \tilde{\bm{y}}^{(h)} \mid \tilde{\bm{X}}^{(h)} \right] \right) \right] .
\label{USE}
\end{align}
This is the sum of squared differences between the expectations for observed data themselves and their estimators.
 In term of the improvement of estimation accuracy for observed data, this loss function may be more natural than before.
 According to the derivation of the conventional $C_p$ criterion, we remove the expectation in the first term, ignore the second term independent of models and asymptotically estimate the third term after setting $\E[\tilde{\bm{y}}^{(h)}\mid \tilde{\bm{X}}^{(h)}]=\tilde{\bm{X}}^{(h)}\bm{\beta}$, and we propose it as a $C_p$ criterion in causal inference.
 In the asymptotic evaluation, a main term is extracted from the contents of the expectation, and we take its expectation explicitly.  
 Then, we denote the criteria derived from \MwSE\ and \MuSE\ by \wcp\ and \ucp, respectively.


\subsection{Criterion for inverse-probability-weighted estimation with known propensity scores}\label{sec3_2}

Let us derive \wcp\ for the inverse-probability-weighted estimation when $\bm{\alpha}$ is known.
 In \eqref{IPWest}, the inversed matrix divided by $N$ is expressed as
\begin{align}
\frac{1}{N} \sum_{h=1}^{H} \sum_{i=1}^{N} \frac{t_{i}^{(h)}}{e_{i}^{(h)}} \bm{X}_i^{(h)\T} \bm{X}_i^{(h)} = \sum_{h=1}^{H} \E \left[ \frac{t^{(h)}}{e^{(h)}} \bm{X}^{(h)\T} \bm{X}^{(h)} \right] \left\{1+\oP(1)\right\} = \bm{I} \left\{1+\oP(1)\right\}.
\label{th1pre1}
\end{align}
The second equality holds because of the assumption for $\bm{X}^{(h)}$ and because the expectation is written as $\E [ \E [ t^{(h)}/e^{(h)} \mid \bm{z} ] \allowbreak \bm{X}^{(h)\T} \bm{X}^{(h)} ] = \E [ \bm{X}^{(h)\T} \bm{X}^{(h)} ]$ from the ignorable treatment assignment condition. 
 In addition, using $\bm{W}^{(h)} (\tilde{\bm{y}} - \tilde{\bm{X}}^{(h)} \bm{\beta}) = \bm{W}^{(h)} (\tilde{\bm{y}}^{(h)} - \tilde{\bm{X}}^{(h)} \bm{\beta}) = \bm{W}^{(h)} \tilde{\bm{\varepsilon}}$, the error of the inverse-probability-weighted estimator is expressed as
\begin{align}
\hat{\bm{\beta}}^{\rm IPW} - \bm{\beta} = \frac{1}{N} \sum_{h=1}^{H} \tilde{\bm{X}}^{(h)\T} \bm{W}^{(h)} \tilde{\bm{\varepsilon}} \left\{1+\oP(1)\right\} = \frac{1}{N} \sum_{h=1}^{H} \sum_{i=1}^{N} \frac{t_{i}^{(h)}}{e_{i}^{(h)}} \bm{X}_i^{(h)\T} \bm{\varepsilon}_i \left\{1+\oP(1)\right\}.
\label{IPWerror1}
\end{align}
Therefore, replacing $\hat{\bm{\beta}}-\bm{\beta}$ with this main term in the third term in the right hand side of \eqref{WSE}, the expectation in it is asymptotically evaluated as
\begin{align}
\E \left[ \tilde{\bm{\varepsilon}}^{\T} \bm{W}^{(h)} \tilde{\bm{X}}^{(h)} \frac{1}{N} \sum_{k=1}^{H} \sum_{j=1}^N \frac{t_j^{(h)}}{e_j^{(h)}} \bm{X}_j^{(h)\T} \bm{\varepsilon}_j \right]
=\frac{1}{N} \sum_{k=1}^{H} \sum_{i,j=1}^{N} \E \left[ \frac{t_{i}^{(h)}}{e_{i}^{(h)}} \bm{\varepsilon}_{i}^{\T} \bm{X}^{(h)}_{i} \frac{t_{j}^{(k)}}{e_{j}^{(k)}} \bm{X}_{j}^{(k)\T} \bm{\varepsilon}_{j} \right].
\label{th1pre2}
\end{align}
This expectation for the case where $i \neq j$ is the product of the expectations for $i$ and $j$ from the independence among samples, and it can be ignored because it holds from the ignorable treatment assignment condition and the independence between  $\bm{X}_i^{(h)}$ and $\bm{\varepsilon}_i$ that 
\begin{align}
\E \left[ \frac{t_{i}^{(h)}}{e_{i}^{(h)}} \bm{\varepsilon}_{i}^{\T} \bm{X}^{(h)}_{i} \right]
= \E \left[ \E \left[ \frac{t_{i}^{(h)}}{e_{i}^{(h)}} \mid \bm{z}_{i} \right] \E [\bm{\varepsilon}_{i} \mid \bm{z}_{i}]^{\T} \bm{X}^{(h)}_{i} \right]
= \E \left[ \bm{\varepsilon}_{i}^{\T} \bm{X}^{(h)}_{i} \right]
= \bm{0}.
\label{th1pre3}
\end{align}
Therefore, we have only to consider the case where $i=j$.
 When $i=j$, we can ignore the expectation for the case of $k\neq h$ because in this case $t_{i}^{(h)}t_{i}^{(k)}=0$, and so \eqref{th1pre2} is expressed as
\begin{align*}
\frac{1}{N} \sum_{i=1}^{N} \E \left[ \frac{t_i^{(h)2}}{e_i^{(h)2}} \bm{\varepsilon}_{i}^{\T} \bm{X}_{i}^{(h)} \bm{X}^{(h)\T}_{i} \bm{\varepsilon}_{i} \right]
= \frac{1}{N} \sum_{i=1}^{N} \E \left[ \frac{1}{e_{i}^{(h)}} \bm{\varepsilon}_{i}^{\T} \bm{X}_{i}^{(h)} \bm{X}^{(h)\T}_{i} \bm{\varepsilon}_{i} \right].
\end{align*}
We obtain this equality from using $t_i^{(h)2}=t_i^{(h)}$ and the ignorable treatment assignment condition similarly to in the derivation of \eqref{th1pre3}.
 Because $(e_{i}^{(h)},\bm{\varepsilon}_{i},\bm{X}_{i}^{(h)})$'s are identically distributed, \wcp\ in the following theorem is derived as a result.
 For the derivation of \ucp, which is also given in the theorem, see Appendix. 

\begin{theorem}{ 
For the case where the propensity score is known, the $C_p$ criteria for the inverse-probability-weighted estimation are given as follows:
\begin{align*}
\wcp = \sum_{h=1}^{H} \left( \tilde{\bm{y}} - \tilde{\bm{X}}^{(h)} \hat{\bm{\beta}}^{\rm IPW} \right)^{\T} \bm{W}^{(h)} \left( \tilde{\bm{y}} - \tilde{\bm{X}}^{(h)} \hat{\bm{\beta}}^{\rm IPW} \right) + 2\sum_{h=1}^{H} \E \left[ \frac{1}{e^{(h)}} \bm{\varepsilon}^{\T} \bm{X}^{(h)} \bm{X}^{(h)\T} \bm{\varepsilon} \right]
\end{align*}
and
\begin{align*}
\ucp = \sum_{h=1}^{H} \left( \tilde{\bm{y}} - \tilde{\bm{X}}^{(h)} \hat{\bm{\beta}}^{\rm IPW} \right)^{\T} \bm{T}^{(h)} \left( \tilde{\bm{y}} - \tilde{\bm{X}}^{(h)} \hat{\bm{\beta}}^{\rm IPW} \right) +2\sigma^2p.
\end{align*}
}
\label{th1}
\end{theorem}

Although the expectation in the penalty term for \wcp\ cannot be calculated in general, we can easily give its consistent estimator such as $\sum_{i=1}^{N} \sum_{h=1}^{H} t^{(h)}_{i} (\bm{y}^{(h)}_{i} - \bm{X}^{(h)}_{i} \hat{\bm{\beta}}^{\rm IPW})^{\T} \bm{X}^{(h)}_{i} \bm{X}^{(h)\T}_{i} (\bm{y}^{(h)}_{i} - \bm{X}^{(h)}_{i} \hat{\bm{\beta}}^{\rm IPW}) / (Ne_i^{(h)2})$. 
 Also in the followings, we propose to use such simple consistent estimators in place of the penalty terms.  

Speaking of the forms of criteria, the penalty term for \QIC$_w$ is the same as for \ucp.
 We can say that the increase of the penalty owing to considering the inverse-probability-weighted estimation and the decrease of the penalty owing to considering the loss function only for observed data are the same amount.
 On the other hand, the goodness-of-fit term for \QIC$_w$ is the same as for \wcp.
 Considering that $2\sum_{h=1}^{H} \E [\bm{\varepsilon}^{\T} \bm{X}^{(h)} \bm{X}^{(h)\T} \bm{\varepsilon}] = 2\sigma^2p$, the penalty in \wcp\ is almost the inversed propensity score times the penalty for \QIC$_w$.
 Thus, we can predict that the performances of \wcp\ and \QIC$_w$ are quite different.


\subsection{Criterion for inverse-probability-weighted estimation with unknown propensity scores}\label{sec3_3}

Let us derive \wcp\ for the inverse-probability-weighted estimation when $\bm{\alpha}$ is unknown.
 As written in Section \ref{sec2_2}, we use the maximum likelihood estimator based on $p(\tilde{\bm{t}} \mid \tilde{\bm{z}}; \bm{\alpha}) = \prod_{i=1}^{N} ( \sum_{h=1}^{H} \allowbreak t_i^{(h)} e_i^{(h)})$ as $\hat{\bm{\alpha}}$.
 Then, letting $\bm{\Lambda}^{(h)} \equiv \E [\bm{X}^{(h)\T}\bm{\varepsilon}(\partial e^{(h)} / \partial \bm{\alpha}^{\T}) / e^{(h)}]$ and $\bm{J} \equiv \sum_{h=1}^H \E [(\partial e^{(h)} / \partial \bm{\alpha}) \allowbreak (\partial e^{(h)} / \partial \bm{\alpha}^{\T}) / e^{(h)}]$, as indicated in \cite{HosKS06}, the error of the inverse-probability-weighted estimator is expressed as
\begin{align}
\hat{\bm{\beta}}^{\rm IPW}-\bm{\beta} = \frac{1}{N} \sum_{h=1}^{H} \sum_{i=1}^{N} \left( \frac{t_{i}^{(h)}}{e_{i}^{(h)}} \bm{X}_i^{(h)\T} \bm{\varepsilon}_i - \bm{\Lambda}^{(h)} \bm{J}^{-1} \sum_{k=1}^{H} \frac{t_{i}^{(k)}}{e_{i}^{(k)}} \frac{\partial e_{i}^{(k)}}{\partial\bm{\alpha}} \right) \{1+\oP(1)\}
\label{IPWerror2}
\end{align}
(see Appendix).
 Using this in the third term in the right hand side of \eqref{WSE}, the expectation is asymptotically evaluated as the expression which is made by adding 
\begin{align*}
-\frac{1}{N} \sum_{k,l=1}^{H} \sum_{i,j=1}^{N} \E \left[ \frac{t_{i}^{(h)}}{e_{i}^{(h)}} \bm{\varepsilon}_i^{\T}  \bm{X}_i^{(h)} \bm{\Lambda}^{(k)} \bm{J}^{-1} \frac{t_{j}^{(l)}}{e_{j}^{(l)}} \frac{\partial e_{j}^{(l)}}{\partial \bm{\alpha}} \right] = -\sum_{k=1}^{H} \E \left[ \bm{\varepsilon}^{\T} \bm{X}^{(h)} \bm{\Lambda}^{(k)} \bm{J}^{-1} \frac{1}{e^{(h)}} \frac{\partial e^{(h)}}{\partial \bm{\alpha}} \right]
\end{align*}
to \eqref{th1pre2}, and then \wcp\ in the following theorem is derived.
 This equality is obtained from the fact that the samples are independently and identically distributed and the ignorable treatment assignment condition.
 See Appendix for more detail, which derives \ucp\ in a similar way.

\begin{theorem}{
For the case where the propensity score is unknown, the $C_p$ criteria for the inverse-probability-weighted estimation are given as follows:
\begin{align*}
\wcp = & \sum_{h=1}^{H} \left( \tilde{\bm{y}} - \tilde{\bm{X}}^{(h)} \hat{\bm{\beta}}^{\rm IPW} \right)^{\T} \bm{W}^{(h)} \left( \tilde{\bm{y}} - \tilde{\bm{X}}^{(h)} \hat{\bm{\beta}}^{\rm IPW} \right) +2\sum_{h=1}^{H} \E \left[ \frac{1}{e^{(h)}} \bm{\varepsilon}^{\T} \bm{X}^{(h)} \bm{X}^{(h)\T} \bm{\varepsilon} \right] 
\\
& -2\sum_{k,h=1}^{H} {\rm tr} \left( \E \left[ \frac{1}{e^{(k)}} \bm{X}^{(k)\T} \bm{\varepsilon} \frac{\partial e^{(k)}}{\partial \bm{\alpha}^{\T}} \right] \E \left[ \sum_{l=1}^H \frac{1}{e^{(l)}} \frac{\partial e^{(l)}}{\partial \bm{\alpha}} \frac{\partial e^{(l)}}{\partial \bm{\alpha}^{\T}} \right]^{-1} \E \left[ \frac{1}{e^{(h)}} \bm{X}^{(h)\T} \bm{\varepsilon}  \frac{\partial e^{(h)}}{\partial \bm{\alpha}^{\T}} \right]^{\T} \right)
\end{align*}
and
\begin{align*}
\ucp = & \sum_{h=1}^{H} \left( \tilde{\bm{y}} - \tilde{\bm{X}}^{(h)} \hat{\bm{\beta}}^{\rm IPW} \right)^{\T} \bm{T}^{(h)} \left( \tilde{\bm{y}} - \tilde{\bm{X}}^{(h)} \hat{\bm{\beta}}^{\rm IPW} \right) + 2\sigma^2p
\\
& -2\sum_{k,h=1}^{H} {\rm tr} \left( \E \left[ \frac{1}{e^{(k)}} \bm{X}^{(k)\T} \bm{\varepsilon} \frac{\partial e^{(k)}}{\partial \bm{\alpha}^{\T}} \right] \E \left[ \sum_{l=1}^H \frac{1}{e^{(l)}} \frac{\partial e^{(l)}}{\partial \bm{\alpha}} \frac{\partial e^{(l)}}{\partial \bm{\alpha}^{\T}} \right]^{-1} \E \left[ \bm{X}^{(h)\T} \bm{\varepsilon}  \frac{\partial e^{(h)}}{\partial \bm{\alpha}^{\T}} \right]^{\T} \right).
\end{align*}
}
\label{th2}
\end{theorem}

From Theorems \ref{th1} and \ref{th2}, it can be seen that the penalty for the unknown propensity score tends to be smaller than that for the known propensity score.
 It is known that the asymptotic variance for the inverse-probability-weighted estimator becomes smaller if the propensity score is estimated even for the case where it is known (see, e.g., \citealt{HenEgu04}).
 The property of the penalties is consistent with this fact.


\subsection{Criterion for doubly robust estimation}\label{sec3_4}

Let us derive \wcp\ for the doubly robust estimation.
 In a similar way in \citet{Hos07}, which derived the asymptotic distribution of the doubly robust estimator for a structural equation model with a missing mechanism, the error of the doubly robust estimator is shown to be expressed as
\begin{align}
\hat{\bm{\beta}}^{\rm DR} - \bm{\beta} = \frac{1}{N} \sum_{h=1}^{H} \sum_{i=1}^N \left\{ \frac{t_{i}^{(h)}}{e_{i}^{(h)}} \bm{X}_i^{(h)\T} \bm{\varepsilon}_i + \left( 1-\frac{t_{i}^{(h)}}{e_{i}^{(h)}} \right) \bm{X}_i^{(h)\T} \E [\bm{\varepsilon}_i \mid \bm{z}_i] \right\} \{1+\oP(1)\}
\label{DRerror}
\end{align}
(see Appendix).
 Its main term does not include the score function for $\bm{\alpha}$, which indicates that $\hat{\bm{\beta}}^{\rm DR}$ is semiparametrically efficient.
 Using it in the third term in the right hand side of \eqref{WSE}, the expectation is asymptotically evaluated as the expression which is made by adding
\begin{align*}
& \frac{1}{N} \sum_{k=1}^{H} \sum_{i,j=1}^{N} \E \left[ \frac{t^{(h)}_{i}}{e_{i}^{(h)}} \bm{\varepsilon}^{\T}_{i} \bm{X}_{i}^{(h)} \left( 1-\frac{t^{(k)}_{j}}{{e_{j}^{(k)}}} \right) \bm{X}^{(k)\T}_{j} \E \left[ \bm{\varepsilon}_{j} \mid \bm{z}_{j} \right] \right] 
\\
& = \sum_{k=1}^{H} \E \left[ \E [\bm{\varepsilon} \mid \bm{z}]^{\T} \bm{X}^{(h)} \bm{X}^{(k)\T} \E [\bm{\varepsilon} \mid \bm{z}] \right] - \E \left[ \frac{1}{e^{(h)}} \E [\bm{\varepsilon} \mid \bm{z}]^{\T} \bm{X}^{(h)} \bm{X}^{(h)\T} \E [\bm{\varepsilon} \mid \bm{z}] \right] 
\end{align*}
to \eqref{th1pre2}, and then \wcp\ in the following theorem is derived.
 This equality is obtained from the fact that the samples are independently and identically distributed and the ignorable treatment assignment condition.
 Note that unlike the case of the inverse-probability-weighted estimation, the expectation does not become $0$ even if $k\neq h$.
 See Appendix for more detail, which derives \ucp\ in a similar way.

\begin{theorem}{
The $C_p$ criteria for the doubly robust estimation are given as follows:
\begin{align*}
\wcp= & \sum_{h=1}^{H} \left( \tilde{\bm{y}} - \tilde{\bm{X}}^{(h)} \hat{\bm{\beta}}^{\rm DR} \right)^{\T}\bm{W}^{(h)} \left( \tilde{\bm{y}} - \tilde{\bm{X}}^{(h)} \hat{\bm{\beta}}^{\rm DR} \right) 
+2\sum_{h=1}^{H} \E \left[ \frac{1}{e^{(h)}} \bm{\varepsilon}^{\T} \bm{X}^{(h)} \bm{X}^{(h)\T}\bm{\varepsilon} \right]
\nonumber \\
& +2\sum_{h,k=1}^{H} \E \left[ \E [\bm{\varepsilon} \mid \bm{z}]^{\T} \bm{X}^{(h)} \bm{X}^{(k)\T} \E [\bm{\varepsilon} \mid \bm{z}] \right]
-2\sum_{h=1}^{H} \E \left[ \frac{1}{e^{(h)}} \E [\bm{\varepsilon} \mid \bm{z}]^{\T} \bm{X}^{(h)} \bm{X}^{(h)\T} \E [\bm{\varepsilon} \mid \bm{z}] \right]
\end{align*}
and
\begin{align*}
\ucp= & \sum_{h=1}^{H} \left( \tilde{\bm{y}} - \tilde{\bm{X}}^{(h)} \hat{\bm{\beta}}^{\rm DR} \right)^{\T} \bm{T}^{(h)} \left( \tilde{\bm{y}} - \tilde{\bm{X}}^{(h)} \hat{\bm{\beta}}^{\rm DR} \right) 
+ 2\sigma^2p
\nonumber \\
&+2\sum_{h,k=1}^{H} \E \left[ e^{(h)} \E[ \bm{\varepsilon} \mid \bm{z}]^{\T} \bm{X}^{(h)} \bm{X}^{(k)\T} \E [\bm{\varepsilon} \mid \bm{z}] \right] 
-2\sum_{h=1}^{H} \E \left[ \E [\bm{\varepsilon} \mid \bm{z}]^{\T} \bm{X}^{(h)} \bm{X}^{(h)\T} \E [\bm{\varepsilon} \mid \bm{z}] \right].
\end{align*}
}
\label{th3}
\end{theorem}


\section{Simulation study}\label{sec4}

\subsection{Setup}\label{sec4_1}

Let us evaluate the performance of the proposed criterion through simulation study using a marginal structural model $y^{(h)}=\sum_{j=0}^pb_{j+1}x^{(h)j}+\varepsilon\ (1\le h\le H)$, which is introduced in Section \ref{sec1}. 
 According to the setting in \cite{PlaBro13}, we set $H=6$.
 In addition, letting $x^{(h)}=h$, we consider a polynomial model whose order $p$ is at most $5$ because $H=6$. 
 As the true structure, let us consider
\begin{align*}
y^{(h)}=1+x^{(h)}+bx^{(h)2}+z_1+\epsilon,
\end{align*}
and we set $b$ is $0.5$, $0.3$ or $0.1$ to examine a second-order polynomial structure which is far from or close to first-order polynomial model.
 We assume that $z_1$ and $\epsilon$ are independently distributed according to a uniform distribution ${\rm U}(-\sqrt{3},\sqrt{3})$ and a Gaussian distribution ${\rm N}(0,1)$, respectively, and then $\varepsilon=z_1+\epsilon$ is a noise with mean $0$ and variance $2$.
 As for the propensity score, letting the true value of $\bm{\alpha}=(\alpha_1,\alpha_2,\alpha_3,\alpha_4,\alpha_5)$ be $(0.8,1.0,0.9,0.7,0.6)$, we assumed that 
\begin{align*}
e^{(h)}\propto \exp (1_{\{h\neq 1\}}\alpha_{h-1}z_1).
\end{align*}
 In addition, we consider $N=100$ or $N=200$ as sample size.

Under this setting, the experiment of selecting $p$ from $\{0,1,\ldots,5\}$ by each criterion is repeated $5000$ times.
 In the $p$-th order polynomial model, using $(p+1) \times (p+1)$ nonsingular matrix  $\bm{A}$ such that $\sum_{h=1}^6 (1, x^{(h)}, \ldots , x^{(h)p})^{\T} (1, x^{(h)}, \ldots, x^{(h)p}) = \bm{A}^{\T} \bm{A}$, we set $\bm{X}^{(h)}=(1, x^{(h)}, \ldots , x^{(h)p}) \bm{A}^{-1}$ and $\bm{\beta} = \bm{A} (b_0, b_1, \ldots, b_p)^{\T}$.
 Then, we can express $y^{(h)} = \bm{X}^{(h)} \bm{\beta} + \varepsilon$ and it holds $\sum_{h=1}^6 \bm{X}^{(h)\T} \bm{X}^{(h)} = \bm{I}_{p+1}$, which enable us to calculate all the $C_p$ criteria.

\subsection{Results}\label{sec4_2}

First, let us investigate whether the asymptotic evaluation approximates the penalty well or not.
 For the third terms in the right hand side of \eqref{WSE} and \eqref{USE}, in Table \ref{tab1}, we compare Monte Carlo evaluations and the asymptotic evaluations in Theorems \ref{th1}, \ref{th2} and \ref{th3}. 
 Note that the asymptotic evaluation in Theorem \ref{th1} for \ucp\ is $2\E[\varepsilon^2]p=2\times 2\times 3=12$ in this setting.
 From the table, we can check that the accuracy of evaluations tends to become high as the sample size increases.
 Considering that the penalty in \QIC$_w$ is $2\E[\varepsilon^2]p=12$, we can say that the penalty in \wcp\ is more than enough close to the Monte Carlo evaluation even if $N=100$.

\begin{table}[t]
\renewcommand{\baselinestretch}{1.2}\selectfont
\def~{\hphantom{0}}
\caption{Monte Carlo and asymptotic evaluations for penalty terms.}
\begin{center}
\begin{tabular}{cccccccccccccc}
 & & & \multicolumn{5}{c}{$N=100$} & & \multicolumn{5}{c}{$N=200$} \\
 & & & \multicolumn{2}{c}{\wcp} & & \multicolumn{2}{c}{\ucp}
 & & \multicolumn{2}{c}{\wcp} & & \multicolumn{2}{c}{\ucp} \\
 & & & MCE & AE & & MCE & AE & & MCE & AE & & MCE & AE \\[5pt]
IPW1 & $b=0.5$
 & & 82.95 & 79.34 & & 12.48 & 12.00 & & 85.88 & 82.09 & & 13.24 & 12.00 \\
 & $b=0.3$
 & & 86.52 & 79.19 & & 13.10 & 12.00 & & 85.30 & 82.73 & & 12.88 & 12.00 \\
 & $b=0.1$
 & & 85.29 & 79.23 & & 13.08 & 12.00 & & 85.77 & 82.67 & & 12.67 & 12.00 \\[5pt]
IPW2 & $b=0.5$
 & & 60.36 & 56.55 & & ~8.87 & 10.90 & & 61.18 & 58.89 & & ~9.21 & ~7.57 \\
 & $b=0.3$
 & & 63.12 & 57.04 & & ~9.02 & 10.89 & & 61.47 & 59.53 & & ~8.87 & 11.13 \\
 & $b=0.1$
 & & 61.61 & 56.91 & & ~9.10 & 10.90 & & 60.82 & 59.35 & & ~8.52 & 11.11 \\[5pt]
DR & $b=0.5$
 & & 50.10 & 49.08 & & ~8.05 & ~7.58 & & 50.50 & 50.40 & & ~8.17 & ~7.83 \\
 & $b=0.3$
 & & 54.11 & 44.52 & & ~7.99 & ~7.56 & & 53.31 & 49.87 & & ~8.06 & ~7.81 \\
 & $b=0.1$
 & & 53.08 & 44.76 & & ~8.05 & ~7.56 & & 51.93 & 49.70 & & ~7.77 & ~7.81 \\
\end{tabular}
\end{center}
\begin{center}
MCE, Monte Carlo evaluation; AE, asymptotic evaluation; IPW1, inverse-probability-weighted estimation with known propensity score; IPW2, inverse-probability-weighted estimation with unknown propensity score; DR, doubly robust estimation.
\end{center}
\label{tab1}
\end{table}


Next, to compare the performances of these criteria, we evaluate the average of $5000$ weighted or unweighted squared errors for the model selected by each criterion.
 In Tables \ref{tab2}, \ref{tab3} and \ref{tab4}, the values are respectively for the inverse-probability-weighted estimation with known propensity scores, for the inverse-probability-weighted estimation with unknown propensity scores and for the doubly robust estimation.
 In all cases, \wcp\ provides clearly smaller squared errors than \QIC$_w$.
 On the other hand, \ucp\ provides larger squared errors than \QIC$_w$ when the true structure is close to the first-order polynomial, while it is sometimes superior to \wcp.
 Thus, basically we propose to use \wcp.

Let us check the selection frequencies of the optimal model, which are given as a reference in the tables.
 Note that, in all tables, the true structure is second-order polynomial.
 When the true structure is extremely close to first-order polynomial, however, it must be appropriate to select the first-order polynomial considering a prediction.
 Therefore, a high selection frequency of the first-order polynomial model does not necessarily indicate an unreasonable model selection.
 Meanwhile, a high selection frequency of more than third-order polynomial is clearly unreasonable.
 In this view point, obviously \QIC$_w$ has a problem.
 On the other hand, we can see that \wcp\ always selects the true second-order polynomial with high frequency.

\begin{table}[t]
\renewcommand{\baselinestretch}{1.2}\selectfont
\def~{\hphantom{0}}
\caption{Average of squared errors and selection frequency for inverse-probability-weighted estimation with known propensity scores.}
\begin{center}
\begin{tabular}{ccccccccccccc}
 & & & & \multicolumn{2}{c}{Average} & & \multicolumn{5}{c}{Selection frequency} \\
 & & & & WSE & USE & & $0$ & $1$ & $2$ & $3$ & $4$ & $5$ \\ [5pt]
 $b=0.5$ & $N=100$
 & QIC$_{w}$ & & 81.02 & 13.67 & & 0.00 & ~0.00 & 10.56 & 12.28 & 22.14 & 55.02 \\
 & & w$C_{p}$ & & 62.55 & 10.61 & & 0.00 & ~0.00 & 68.52 & 14.40 & ~9.24 & ~7.84 \\ 
 & & u$C_{p}$ & & 59.38 & 10.15 & & 0.00 & ~0.00 & 66.32 & 18.70 & ~8.42 & ~6.56 \\[5pt] 
 & $N=200$
 & QIC$_{w}$ & & 82.42 & 13.92 & & 0.00 & ~0.00 & 11.16 & 12.86 & 22.86 & 53.12 \\
 & & w$C_{p}$ & & 62.35 & 10.62 & & 0.00 & ~0.00 & 72.20 & 13.92 & ~7.78 & ~6.10 \\
 & & u$C_{p}$ & & 59.59 & 10.22 & & 0.00 & ~0.00 & 64.86 & 21.98 & ~7.30 & ~5.86 \\[5pt]
 $b=0.3$ & $N=100$
 & QIC$_{w}$ & & 82.18 & 13.92 & & 0.00 & ~0.00 & 11.72 & 12.02 & 21.52 & 54.74 \\
 & & w$C_{p}$ & & 63.84 & 10.87 & & 0.00 & ~0.02 & 69.64 & 13.20 & ~8.74 & ~8.40. \\ 
 & & u$C_{p}$ & & 81.62 & 14.11 & & 0.00 & ~7.36 & 60.36 & 17.08 & ~7.72 & ~7.48 \\[5pt] 
 & $N=200$
 & QIC$_{w}$ & & 82.55 & 13.93 & & 0.00 & ~0.00 & 11.72 & 12.76 & 22.16 & 53.36 \\
 & & w$C_{p}$ & & 62.57 & 10.63 & & 0.00 & ~0.00 & 71.72 & 13.72 & ~8.16 & ~6.40 \\
 & & u$C_{p}$ & & 68.42 & 11.74 & & 0.00 & ~1.54 & 64.70 & 20.28 & ~7.42 & ~6.06 \\[5pt]
 $b=0.1$ & $N=100$
 & QIC$_{w}$ & & 82.03 & 13.91 & & 0.00 & ~2.34 & ~9.76 & 11.86 & 22.38 & 53.66 \\
 & & w$C_{p}$ & & 66.21 & 11.35 & & 0.00 & 31.70 & 41.92 & 12.28 & ~7.24 & ~6.86 \\ 
 & & u$C_{p}$ & & 72.63 & 12.84 & & 0.00 & 87.06 & ~2.36 & ~5.38 & ~2.60 & ~2.60 \\[5pt] 
 & $N=200$
 & QIC$_{w}$ & & 82.52 & 13.91 & & 0.00 & ~0.72 & 11.28 & 12.54 & 22.08 & 53.38 \\
 & & w$C_{p}$ & & 67.02 & 11.42 & & 0.00 & 14.44 & 59.32 & 12.68 & ~7.30 & ~6.26 \\
 & & u$C_{p}$ & & 104.41 & 18.22 & & 0.00 & 93.96 & ~0.42 & ~3.32 & ~1.04 & ~1.26 \\
\end{tabular}
\end{center}
\begin{center}
WSE, weighted squared error; USE, unweighted squared error.
\end{center}
\label{tab2}
\end{table}

\begin{table}[t]
\renewcommand{\baselinestretch}{1.2}\selectfont
\def~{\hphantom{0}}
\caption{Average of squared errors and selection frequency for inverse-probability-weighted estimation with unknown propensity scores.}
\begin{center}
\begin{tabular}{ccccccccccccc}
  & & & & \multicolumn{2}{c}{Average} & & \multicolumn{5}{c}{Selection frequency} \\
 & & & & WSE & USE & & $0$ & $1$ & $2$ & $3$ & $4$ & $5$ \\[5pt] 
 $b=0.5$ & $N=100$
 & QIC$_{w}$ & & 57.17 & ~9.91 & & 0.00 & ~0.00 & 20.70 & 15.28 & 21.12 & 42.90 \\
 & & w$C_{p}$ & & 48.10 & ~8.37 & & 0.00 & ~0.00 & 63.44 & 15.12 & 11.18 & 10.26 \\ 
 & & u$C_{p}$ & & 41.45 & ~7.26 & & 0.00 & ~0.02 & 73.96 & 15.90 & ~6.28 & ~3.84 \\[5pt] 
 & $N=200$
 & QIC$_{w}$ & & 53.51 & ~9.22 & & 0.00 & ~0.00 & 22.72 & 16.60 & 21.70 & 38.98 \\
 & & w$C_{p}$ & & 43.74 & ~7.57 & & 0.00 & ~0.00 & 70.02 & 14.32 & ~8.96 & ~6.70 \\
 & & u$C_{p}$ & & 38.66 & ~6.74 & & 0.00 & ~0.00 & 72.18 & 19.80 & ~5.12 & ~2.90 \\[5pt]
 $b=0.3$ & $N=100$
 & QIC$_{w}$ & & 58.14 & 10.10 & & 0.00 & ~0.00 & 20.86 & 15.16 & 21.30 & 42.68 \\
 & & w$C_{p}$ & & 49.30 & ~8.59 & & 0.00 & ~0.00 & 62.98 & 15.14 & 10.36 & 11.52 \\ 
 & & u$C_{p}$ & & 85.74 & 15.41 & & 0.00 & 14.72 & 62.24 & 14.28 & ~4.98 & ~3.78 \\[5pt]
 & $N=200$
 & QIC$_{w}$ & & 54.34 & ~9.34 & & 0.00 & ~0.00 & 23.20 & 16.56 & 20.40 & 39.84 \\
 & & w$C_{p}$ & & 44.43 & ~7.67 & & 0.00 & ~0.00 & 69.98 & 14.28 & ~8.62 & ~7.12 \\
 & & u$C_{p}$ & & 81.28 & 14.42 & & 0.00 & ~6.72 & 67.04 & 17.44 & ~5.38 & ~3.42 \\[5pt]
 $b=0.1$ & $N=100$
 & QIC$_{w}$ & & 57.72 & 10.05 & & 0.00 & ~2.46 & 19.34 & 15.46 & 20.68 & 42.06 \\
 & & w$C_{p}$ & & 49.84 & ~8.83 & & 0.00 & 16.34 & 49.54 & 14.08 & ~9.90 & 10.14 \\ 
 & & u$C_{p}$ & & 58.96 & 11.08 & & 0.00 & 89.50 & ~3.56 & ~3.82 & ~1.70 & ~1.42 \\[5pt] 
 & $N=200$
 & QIC$_{w}$ & & 53.59 & ~9.21 & & 0.00 & ~0.52 & 22.88 & 16.60 & 20.24 & 39.76 \\
 & & w$C_{p}$ & & 44.72 & ~7.75 & & 0.00 & ~3.74 & 67.10 & 13.40 & ~8.04 & ~7.72 \\
 & & u$C_{p}$ & & 91.43 & 16.64 & & 0.00 & 95.14 & ~1.34 & ~2.34 & ~0.78 & ~0.40 \\
\end{tabular}
\end{center}
\begin{center}
WSE, weighted squared error; USE, unweighted squared error.
\end{center}
\label{tab3}
\end{table}

\begin{table}[t]
\renewcommand{\baselinestretch}{1.2}\selectfont
\def~{\hphantom{0}}
\caption{Average of squared errors and selection frequency for doubly robust estimation.}
\begin{center}
\begin{tabular}{ccccccccccccc}
 & & & & \multicolumn{2}{c}{Average} & & \multicolumn{5}{c}{Selection frequency} \\
 & & & & WSE & USE & & $0$ & $1$ & $2$ & $3$ & $4$ & $5$ \\[5pt] 
 $b=0.5$ & $N=100$
 & QIC$_{w}$ & & 45.83 & ~7.80 & & 0.00 & ~0.00 & 28.44 & 17.04 & 20.46 & 34.06 \\
 & & w$C_{p}$ & & 40.10 & ~6.86 & & 0.00 & ~0.00 & 58.90 & 15.54 & 12.38 & 13.18 \\ 
 & & u$C_{p}$ & & 39.09 & ~6.69 & & 0.00 & ~0.10 & 56.82 & 18.98 & 12.20 & 11.90 \\[5pt] 
 & $N=200$
 & QIC$_{w}$ & & 44.13 & ~7.46 & & 0.00 & ~0.00 & 28.84 & 16.86 & 21.24 & 33.06 \\
 & & w$C_{p}$ & & 36.94 & ~6.27 & & 0.00 & ~0.00 & 67.56 & 15.14 & ~9.28 & ~8.02 \\
 & & u$C_{p}$ & & 35.74 & ~6.07 & & 0.00 & ~0.00 & 58.36 & 21.64 & ~9.84 & 10.16 \\[5pt]
 $b=0.3$ & $N=100$
 & QIC$_{w}$ & & 46.27 & 7.88 & & 0.00 & ~0.00 & 28.60 & 16.70 & 19.46 & 35.24 \\
 & & w$C_{p}$ & & 40.74 & ~6.96 & & 0.00 & ~0.02 & 58.22 & 15.76 & 11.88 & 14.12 \\ 
 & & u$C_{p}$ & & 87.52 & 15.37 & & 0.00 & 16.16 & 46.98 & 15.96 & 10.22 & 10.68 \\[5pt]
 & $N=200$
 & QIC$_{w}$ & & 45.23 & ~7.63 & & 0.00 & ~0.00 & 29.36 & 17.54 & 19.22 & 33.88 \\
 & & w$C_{p}$ & & 38.42 & ~6.51 & & 0.00 & ~0.00 & 67.18 & 14.78 & ~9.20 & ~8.84 \\
 & & u$C_{p}$ & & 81.20 & 14.15 & & 0.00 & ~7.04 & 54.04 & 19.20 & 10.38 & ~9.34 \\[5pt]
 $b=0.1$ & $N=100$
 & QIC$_{w}$ & & 47.52 & ~8.09 & & 0.00 & ~4.90 & 24.68 & 16.42 & 19.50 & 34.50 \\
 & & w$C_{p}$ & & 43.61 & ~7.46 & & 0.00 & 13.32 & 47.12 & 14.74 & 10.80 & 14.02 \\ 
 & & u$C_{p}$ & & 58.60 & 10.14 & & 0.00 & 82.54 & ~4.28 & ~4.78 & ~3.60 & ~4.80 \\[5pt] 
 & $N=200$
 & QIC$_{w}$ & & 44.60 & ~7.54 & & 0.00 & ~0.82 & 28.96 & 17.16 & 19.98 & 33.08 \\
 & & w$C_{p}$ & & 38.54 & ~6.55 & & 0.00 & ~3.28 & 65.88 & 14.80 & ~8.20 & ~7.84 \\
 & & u$C_{p}$ & & 92.47 & 15.93 & & 0.00 & 92.90 & ~1.68 & ~2.68 & ~1.16 & ~1.58 \\
\end{tabular}
\end{center}
\begin{center}
WSE, weighted squared error; USE, unweighted squared error.
\end{center}
\label{tab4}
\end{table}



The doubly robust estimator has semiparametric efficiency when both the conditional expectation and the propensity score are correctly specified, and we derive \wcp\ and \ucp\ in Section \ref{sec3_4} under this condition.
 On the other hand, this estimator is consistent even if either of them is misspecified, and this is its remarkable property.
 Then, we investigate the behaviors of the criteria under misspecification as a sensitivity analysis in Appendix.
 We can see that \wcp\ provides clearly smaller squared errors than \QIC$_w$ also in this case and that the values of selection frequency for \wcp\ are similar to in Table \ref{tab4}.


\section{Data analysis}\label{sec5}


\section{Discussion}\label{sec6}

\subsection{Modification of the risk function}\label{sec6_1}

In this paper, for the marginal structural model, which plays an important role in causal inference, we have considered two kinds of mean squared errors peculiar to this type of causal inference, and information criteria \ucp\ and \wcp\ have been derived as their asymptotically unbiased estimators.
 In addition, through simulation studies, we have shown that \wcp\ always performs well although it is occasionally inferior to \ucp, more concretely speaking, \wcp\ is clearly superior to \QIC$_w$ in terms of the mean squared error and the selection frequency.

While the mean squared error which \wcp\ is based on is naturally considered, its improvement is an important future theme. 
 Here, as its first step, we consider to modify the expectation in the definition of the mean squared error. 
 In realty, this expectation is taken to make the evaluation of the squared error possible, and it can do no better than evaluate it without taking the expectation if possible (see, e.g., \citealt{Efr86}).
 Therefore, it is desirable to take a conditional expectation which does not lose information of data and which can be evaluated explicitly.
 For example, \cite{VaiBla05} proposes a conditional AIC for mixed models by considering a conditional expectation of a loss function given the random coefficients.

Let $a^{(h)}$ be a value of the estimate of the propensity score $e^{(h)}$ based on finite real samples.
 Then, let us condition that the frequency of being $t^{(h)}=1$ in samples whose estimate of $e^{(h)}$ is the value is kept to be $a^{(h)}$ also in the asymptotics.
 Specifically, letting $\hat{e}_i^{(h)}$ be the estimate of $e_i^{(h)}$, and letting $A^{(h)} \equiv \{i \mid \hat{e}_i^{(h)} = a^{(h)}\}$, we condition an event that it holds $|A^{(h)}| a^{(h)} - 1/2 < \sum_{i\in A^{(h)}} t_i^{(h)} \le |A^{(h)}| a^{(h)} + 1/2$ for any $a^{(h)}$.
 That is, denoting this event by $B$, we consider 
\begin{align*}
\sum_{h=1}^{H} \E \left[ \left( \tilde{\bm{y}} - \tilde{\bm{X}}^{(h)} \bm{\beta} \right)^{\T} \bm{W}^{(h)}\tilde{\bm{X}}^{(h)} \left( \hat{\bm{\beta}}^{\rm IPW} - \bm{\beta} \right) \mid B \right]
\end{align*}
as a penalty in \wcp\ for the inverse-probability-weighted estimation with the unknown propensity score.
 Under this condition, asymptotically $(t_i^{(1)},\ldots,t_i^{(H)})$'s are regarded as independent samples from multinomial distribution ${\rm Mn}(1,(\hat{e}_i^{(1)},\ldots,\hat{e}_i^{(H)}))$, and so we obtain
\begin{align*}
\sum_{h=1}^{H} \left( \tilde{\bm{y}} - \tilde{\bm{X}}^{(h)} \hat{\bm{\beta}}^{\rm IPW} \right)^{\T} \bm{W}^{(h)} \left( \tilde{\bm{y}} - \tilde{\bm{X}}^{(h)} \hat{\bm{\beta}}^{\rm IPW} \right) + 2\sum_{h=1}^{H} \E \left[ \frac{1}{\hat{e}^{(h)}} \bm{\varepsilon}^{\T} \bm{X}^{(h)} \bm{X}^{(h)\T} \bm{\varepsilon} \right]
\end{align*}
as a $C_p$ criterion similarly to in Section \ref{sec3}.
 In simulation study for six kinds of $(b,N)$ in Table \ref{tab3}, this criterion reduce $6.36$ in the mean squared error and increase $15.6\%$ in the selection frequency of the true second-order polynomial on average in comparison with \wcp\ in Theorem \ref{th2}.
 We can say that this idea has a potential for an improvement of the criterion.


\subsection{Application to missing data analysis}\label{sec6_2}

The marginal structural model attracts attention especially in medical and epidemiological statistics, and the model itself and estimation method for it are being developed rapidly.
 However, there is no information criterion obtained according to its classical derivation for this model even in the simplest setting. 
 This is the reason why we restrict our setting to be simple, and to customize our criterion for more realistic problem is an important future theme.
 The examples are to customize it for a model with time-dependent covariates (\citealt{BanRob05}), for structural equation model in causal inference (\citealt{HosKS06}, \citealt{Hos07}), for multiple robust estimation (\citealt{HanWan13}), for targeting the average treatment effect on the treated (\citealt{SatMat03}).
 As one of the easiest examples, here we customize our criterion for missing data analysis (\citealt{Rub85}, \citealt{RobRZ94}). 
 To avoid redundant statements, we treat only \wcp\ like in Section \ref{sec6_1}.

Let us consider a model $\bm{y}=\bm{X}\bm{\beta}+\bm{\varepsilon}$, and we assume that the outcome $\bm{y}$ is observed or unobserved when a missing indicator $t$ is $1$ or $0$, respectively.
 Here, $\bm{X}$ is an independent variable matrix satisfying $\E [\bm{X}^{\T} \bm{X}] = \bm{I}$, $\bm{\varepsilon}$ is an error vector with mean $\bm{0}$, and we suppose that a confounder vector $\bm{z}$ between $t$ and $\bm{\varepsilon}$ is observed while they are correlated.
 In addition, we assume a missing at random condition $\bm{y} \perp t \mid \bm{z}$. 

We have $N$ independent samples from this model, and as before, we put ${}_i$ like $\bm{y}_i$ in variables for the $i$-th sample and $\tilde{\ }$ like $\tilde{\bm{y}}$ in vectors and matrices made by gathering variables for the $N$ samples.
 Letting $\bm{W} = {\rm diag} (\bm{I}_{r} / e_i)$, where $e_i = \P (t_i=1 \mid \bm{z}_i; \bm{\alpha})$ is the propensity score, the inverse-probability-weighted estimator is given by removing the expressions with respect to $h$, $\sum_{h=1}^H$ and ${}^{(h)}$, in \eqref{IPWest}, that is, 
\begin{align*}
\hat{\bm{\beta}}^{\rm IPW} = \left( \tilde{\bm{X}}^{\T} \bm{W} \tilde{\bm{X}} \right)^{-1} \tilde{\bm{X}}^{\T} \bm{W} \tilde{\bm{y}}.
\end{align*}
Then, its error is given by removing the expressions with respect to $h$ in \eqref{IPWerror1}, and we can derive 
\begin{align*}
\wcp = (\tilde{\bm{y}} - \tilde{\bm{X}} \hat{\bm{\beta}}^{\rm IPW})^{\T} \bm{W} (\tilde{\bm{y}} - \tilde{\bm{X}} \hat{\bm{\beta}}^{\rm IPW}) + 2\E \left[ \frac{1}{e} \bm{\varepsilon}^{\T} \bm{X} \bm{X}^{\T} \bm{\varepsilon} \right]
\end{align*}
as a $C_p$ criterion for the inverse-probability-weighted estimation with known $\bm{\alpha}$ by defining \MwSE\ similarly to in \eqref{WSE}.
 When $\bm{\alpha}$ is unknown, we use the maximum likelihood estimator $\hat{\bm{\alpha}}$ based on $\P (\tilde{\bm{t}} \mid \tilde{\bm{z}}; \bm{\alpha}) = \prod_{i=1}^N \{ t_i e_i + (1-t_i) (1-e_i) \}$.
Then, letting $\bm{\Lambda} \equiv \E [\bm{X}^{\T} \bm{\varepsilon} (\partial e / \partial \bm{\alpha}^{\T}) / e]$ and $\bm{J} \equiv \E [(\partial e / \partial \bm{\alpha}) (\partial e / \partial \bm{\alpha}^{\T}) / e]$, the error of the inverse-probability-weighted estimator is expressed as
\begin{align*}
\hat{\bm{\beta}}^{\rm IPW}-\bm{\beta} = \frac{1}{N} \sum_{i=1}^{N} \left\{ \frac{t_i}{e_i} \bm{X}_i^{\T} \bm{\varepsilon}_i - \bm{\Lambda} \bm{J}^{-1} \left( \frac{t_i}{e_i} - \frac{1-t_i}{1-e_i} \right) \frac{\partial e_i}{\partial\bm{\alpha}} \right\} \{1+\oP(1)\}.
\end{align*}
Although this is not given by simply removing the expressions with respect to $h$ and $k$ in \eqref{IPWerror2}, \wcp\ is given by removing them in \wcp\ in Theorem \ref{th2}, that is, the penalty term becomes
\begin{align*}
2\E \left[ \frac{1}{e} \bm{\varepsilon}^{\T} \bm{X} \bm{X}^{\T} \bm{\varepsilon} \right] - 2{\rm tr} \left( \E \left[ \frac{1}{e} \bm{X}^{\T} \bm{\varepsilon} \frac{\partial e}{\partial \bm{\alpha}^{\T}} \right] \E \left[ \frac{1}{e} \frac{\partial e}{\partial \bm{\alpha}} \frac{\partial e}{\partial \bm{\alpha}^{\T}} \right]^{-1} \E \left[ \frac{1}{e} \bm{X}^{\T} \bm{\varepsilon}  \frac{\partial e}{\partial \bm{\alpha}^{\T}} \right]^{\T} \right).
\end{align*}
For the doubly robust estimator
\begin{align*}
\hat{\bm{\beta}}^{\rm DR} = \left( \tilde{\bm{X}}^{\T} \tilde{\bm{X}} \right)^{-1} \left\{ \tilde{\bm{X}}^{\T} \bm{W} \tilde{\bm{y}} + \tilde{\bm{X}}^{\T} \left( \bm{I}-\bm{W} \right) \E \left[ \tilde{\bm{y}} \mid \tilde{\bm{z}}; \bm{\gamma} \right] \right\},
\end{align*}
the error is given by simply removing the expressions with respect to $h$ in \eqref{DRerror}, and then the penalty term in \wcp\ becomes
\begin{align*}
2\E \left[ \frac{1}{e} \bm{\varepsilon}^{\T} \bm{X} \bm{X}^{\T} \bm{\varepsilon} \right] + 2\E \left[ \left( 1-\frac{1}{e} \right) \E [\bm{\varepsilon} \mid \bm{z}]^{\T} \bm{X} \bm{X}^{\T} \E [\bm{\varepsilon} \mid \bm{z}] \right].
\end{align*}


\section*{Appendix}



\bibliography{List}

\end{document}